\documentclass{article}
\usepackage{graphicx} % Required for inserting images
\usepackage{siunitx}
\usepackage{natbib}
\usepackage{authblk}

\title{Filling the Gap: Atom Probe Tomography of Porous Structures Enabled by Site-Specific SEMGlu™ Curing}
\author[1]{Lukas Worch}
\author[1]{James Douglas}
\author[2]{Kavin Arunasalam}
\author[1,3,4]{Baptiste Gault}
\author[2]{Valeria Nicolosi}
\author[1,*]{Michele Shelly Conroy}
\affil[1]{Department of Materials, Imperial College London, South Kensington Campus, SW7 2AZ, United Kingdom}
\affil[2]{School of Chemistry, CRANN and AMBER Research Centres, Trinity College Dublin, Dublin 2, Ireland}
\affil[3]{Max Planck Institute for Sustainable Materials, Max-Planck-Str. 1, 40237, D\"usseldorf, Germany}
\affil[4]{now at Univ Rouen Normandie, INSA Rouen Normandie, CNRS, Groupe de Physique des Materiaux, UMR 6634,
F-76000, Rouen, France}
\affil[*]{Corresponding Author. mconroy@imperial.ac.uk}
\date{}

\begin{document}

\maketitle

\section*{Abstract}
Porous microstructures, while central to many functional materials, remain difficult to characterize quantitatively by atom probe tomography (APT). Although several strategies have been proposed in the past, most remain constrained by significant practical or technical limitations. Here, we introduce an \textit{in situ} pore-filling approach that integrates seamlessly into conventional specimen preparation workflows. The method employs a vacuum-compatible resin that is rapidly cured by the electron beam during ion-beam–based preparation, eliminating the need for additional instrumentation or extensive sample handling. We demonstrate the effectiveness of this approach using a porous SnSe + MXene battery electrode, a material system otherwise difficult to analyze via APT characterisation. This method offers a robust, accessible solution for extending APT analysis to porous materials.

% \boxedtext{
% \begin{itemize}
% \item Key boxed text here.
% \item Key boxed text here.
% \item Key boxed text here.
% \end{itemize}}

\maketitle

\section*{Introduction}

Atom probe tomography (APT) is capable of three-dimensional compositional mapping with nanometer-scale resolution and parts-per-million elemental sensitivity across all elements \citep{kelly_atom_2007, kelly_atom_2012, gault_atom_2021}. However, applying APT to porous materials presents unique challenges. First, the specimen geometry itself further amplifies these difficulties. Because APT specimens must be sharpened to a needle-like shape with a tip radius on the order of 100~nm, the presence of pores of comparable size makes specimen preparation particularly challenging.

Second, when the field evaporation front encounters a void, the absence of mechanical support and the concentration of the electric field at void edges can trigger instability in the field evaporation process that cause abrupt changes in the detection rate. In extreme cases, large pores or clusters of voids may  compromise the specimen’s structural integrity during analysis causing  catastrophic partial of complete fracture of the specimen 
\citep{pfeiffer_characterization_2015}.

As they open to the vacuum, voids and pores create a highly heterogeneous electrostatic field distribution that cause both local variations in the evaporation rate as well as severe trajectory aberrations and overlaps \citep{larson2013atom, vurpillot_reconstructing_2013} that lead to severe point density fluctuations within the reconstructed point clouds \citep{wang_interpreting_2020, gault2023transmission}. These effects manifest as artifacts such as apparent compositional gradients, and point density fluctuations that can mimic real chemical heterogeneity \citep{miller2011detecting, larson2015encapsulation, wang_interpreting_2020, gault2023transmission}. Consequently, interpreting composition and morphology near voids becomes highly uncertain unless the evaporation process is modeled or validated through complementary computational or experimental microscopy techniques \citep{wang_interpreting_2020, dubosq_analysis_2020}.

These findings collectively highlight the critical need to account for pore-induced effects in both data reconstruction and specimen preparation when analyzing porous materials by APT.

The last decade has seen the development of several different methods to address this problem. For APT sample preparation taking place in a focused-ion beam (FIB), it is possible to fill pores directly with use of the gas injection system (GIS) injecting a precursor gas and curing it with the electron or ion beam. This is most commonly a platinum precursor gas, though others are also available to deposit e.g. carbon or tungsten. Previous attempts at this method \citep{pfeiffer_characterization_2015, barroo_aggregated_2020} have shown it to be very effective, though also quite time consuming as the penetration of the precursor gas into the pore network is typically not complete and filling needs to be repeated several times throughout the thinning process, though with the distinct advantage of not requiring any further materials or instruments than those already used for specimen preparation. The carbon content of the precursor gases and final deposited product also varies significantly from metal to metal, and the final measured composition is additionally affected both deposition and analysis conditions, making a systematic approach to analysis of these systems difficult \citep{diercks_electron_2017, perea_environmental_2017}. 

Another example is electrodeposition, whereby various metals can be deposited directly into the pore structures of materials, allowing for freedom to select for compatibility with the material being analysed \citep{el-zoka_enhanced_2017, el-zoka_nanoscale_2018, kim_new_2018}. However, these processes do not work on insulating materials without first depositing an intermediate layer \citep{obinata_direct_2025}. For the imaging of nanoparticles specifically, use of methods such as sputter-coating \citep{felfer_revealing_2014}, electron beam assisted chemical vapour deposition \citep{felfer_new_2015}, and encapsulation in organic resin \citep{perea_atom_2016} have all been reported. While these methods are effective, they are not always neatly transferable to bulk porous samples, either because they are designed from the ground up for nanoparticles, or because percolation of the filling material becomes difficult with bulk samples. 

More recently, use of an alloy known as Field's metal has been shown to be effective for filling of pores with liquid metal, that then hardens and can be processed further \citep{kim_liquid_2022}. This method is potent and versatile, but requires heating of the sample to around \SI{70}{\celsius}, which could potentially damage temperature-sensitive samples.

In most cases, these methods are applied to bulk samples, with further sample preparation occurring afterwards, and each  covers only specific use cases. The development of broadly applicable methods of pore filling hence remains critical. 

Here, we present a method of pore-filling via electron-beam cured resin, in the form of SEMGlu\texttrademark, originally designed as a E-beam activated adhesive. The product is commercially available, easy to handle and cure, fully vacuum compatible, and cures into a conductive material when exposed to the electron beam \citep{kleindiek_nanotechnik_gmbh_semglu_nodate, luysberg_e-beam_2008, rout_adhesive-based_2018}. When cured, the strength of the glue is comparable to other epoxy-based resins \citep{rummel_achieving_2009}, and should thus offer stability in line with previously demonstrated resin-based APT preparation methods, such as that demonstrated by \citet{perea_atom_2016}. Pore filling is performed via insertion of a partially completed sample into the glue. The sample is thinned to an extent that the relevant pores are part of a continuous network to the outside of the lamella to allow for percolation of the glue. While full insertion into the glue is possible, it also shows a prominent capillary effect, capable of drawing the material into the relevant pores. As the glue was designed from the ground up for the scanning electron microscope (SEM), this procedure can be performed in-situ after lamella liftout, and the glue then cured with either the electron beam, or through secondary electrons generated during ion beam imaging. The lamella can then be further processed into an APT specimen through existing approaches. The result is a fast and efficient specimen preparation method that significantly increases possibilities for analysis of porous samples in APT, especially in cases of closed pore networks.

\section*{Materials \& Methods}

\subsection*{Sample Preparation}

APT specimen preparation was completed on a Scios Dualbeam Ga FIB-SEM from Thermo Fisher Scientific using standard FIB site-specific methods \citep{thompson_situ_2007}. Imaging of uncured glue was done at low beam currents ($<$\SI{100}{\pico\ampere}), which will not cure the glue \citep{kleindiek_nanotechnik_gmbh_semglu_nodate}. Decrease of voltage may also be useful to avoid curing, though it was found that low currents are sufficient even at \SI{30}{\kilo\volt}. Curing was completed at higher currents of \SI{1}{\nano\ampere} or above.

\subsection*{Atom Probe Tomography}

APT measurements were conducted using a Local Electrode Atom Probe (LEAP) 5000 XR from Cameca. Specimens were analysed in laser pulsing mode, with a pulse energy of \SI{30}{\pico\joule}, a base temperature of \SI{50}{\kelvin}, and an average detection rate of 5 ions per 1000 pulses. The pulsing rate was automatically adjusted to detect heavy Sn and Se compounds that occur between 50 and 250 Da. Reconstructions were created in IVAS 6.3.1 using fixed shank angles.

\subsection*{Materials}

SEMGlu\texttrademark\: is an electron-cured and vacuum compatible resin adhesive manufactured by Kleindiek Nanotechnik GmbH. Samples were mounted on a 22 post microarray (23265) manufactured by Cameca.

The pore filling process was demonstrated on SnSe nanoparticles inside a MXene supporting structure, a model structure of academic and industrial relevance. In recent years, MXene based structures have become an extremely promising and well studied area of energy storage research\citep{li_3d_2020, xu_progress_2021, chen_mxene_2024}. As such, they have been investigated with a number of techniques, among them APT \citep{kramer_nearatomicscale_2024}. However, the inherently porous nature of 3D structures made from layered 2D materials also makes them extremely difficult to measure in APT without additional preparation steps that involve filling these pores.

SnSe particles were created from commercially available crystals (Alfa Aesar), ground into powder and then sonicated in isopropyl alcohol. MXenes were synthesized from MAX precursors via known HF etching methods \citep{lim_fundamentals_2022}. MXene and SnSe were combined by magnetic stirring overnight, and then cast and dried on copper foil in layers of \SI{10}{\micro\metre}.

\section*{Results \& Discussion}

\begin{figure}[h]
    \centering
    \includegraphics[width=0.9\linewidth]{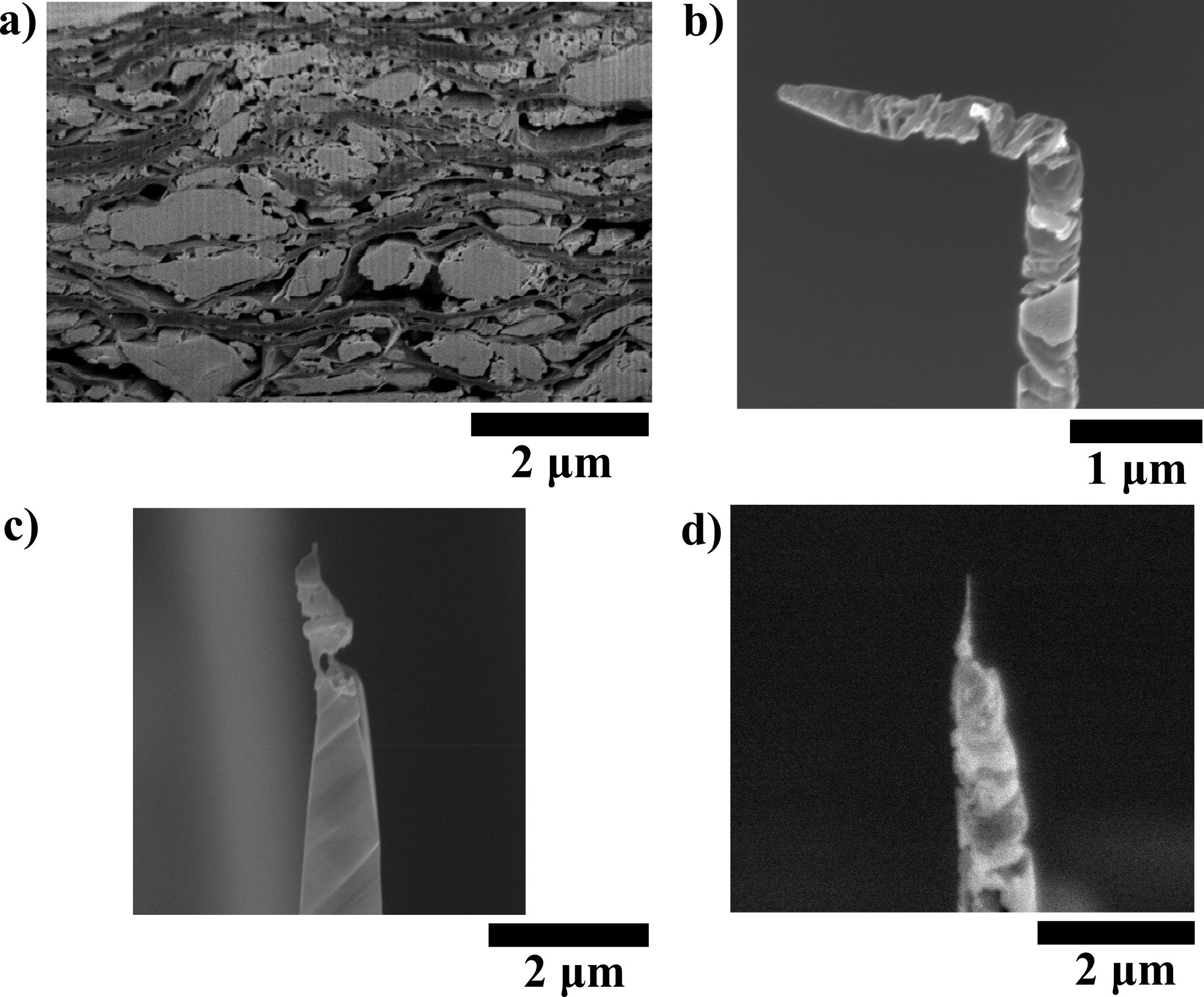}
    \caption{SEM images of SnSe electrode material showing a) a cross-section of the material with SnSe nanoparticles (light contrast) surrounded by MXene sheets (dark contrast), and b-d) APT specimens milled from the material, showing damage to the integrity of the specimens due to the porous structure}
    \label{fig:SnSePorosity}
\end{figure}

The preparation of SnSe particles with MXene sheets as binding material produces electrodes with high porosity. An SEM image of a cross-section of an SnSe electrode can be seen in Figure~\ref{fig:SnSePorosity}a). The pores vary in size significantly, with the largest up to around \SI{1}{\micro\metre} in diameter, and the smallest around \SI{10}{\nano\metre}. Due to a combination of the size variation in the pores, the sheet-like nature of the MXenes winding through the material, and the fact that some of the pores are fully enclosed within the particles or sheets, accessing the entire pore network in the bulk material is nearly impossible.

Nonetheless, the presence of the pores is a significant hindrance to APT analysis. While the material can be mechanically stable in bulk, the number of pores and their size means that once lifted out and thinned to a needle shape, APT specimens become very mechanically weak. This often manifests itself as a bending of the specimen in some part of the shank, as can be seen in Figure~\ref{fig:SnSePorosity}b). This renders the specimen unsuitable for analysis, and necessitates either abandonment of the specimen entirely, or removal of the bent section so that a section further down may be thinned, both of which come with significant loss of time.

Even in cases where a specimen can be fabricated, the connection between the portion of the sample to be analyzed and the rest of the shank is often compromised. For instance, one of the pores may interrupt this connection, as seen in Figure~\ref{fig:SnSePorosity}c), or the top of the specimen may be very thin and therefore susceptible to fracture, as shown in Figure~\ref{fig:SnSePorosity}d). In both situations, the specimens are vulnerable to damage during transfer to the LEAP, or, if an analysis can be initiated, are prone to premature fracture that yields only limited data. It should be noted that most specimen fractures in APT are related to mechanical failure of the specimen~\cite{wilkes1972fracture}.

\begin{figure}[h]
    \centering
    \includegraphics[width=0.9\linewidth]{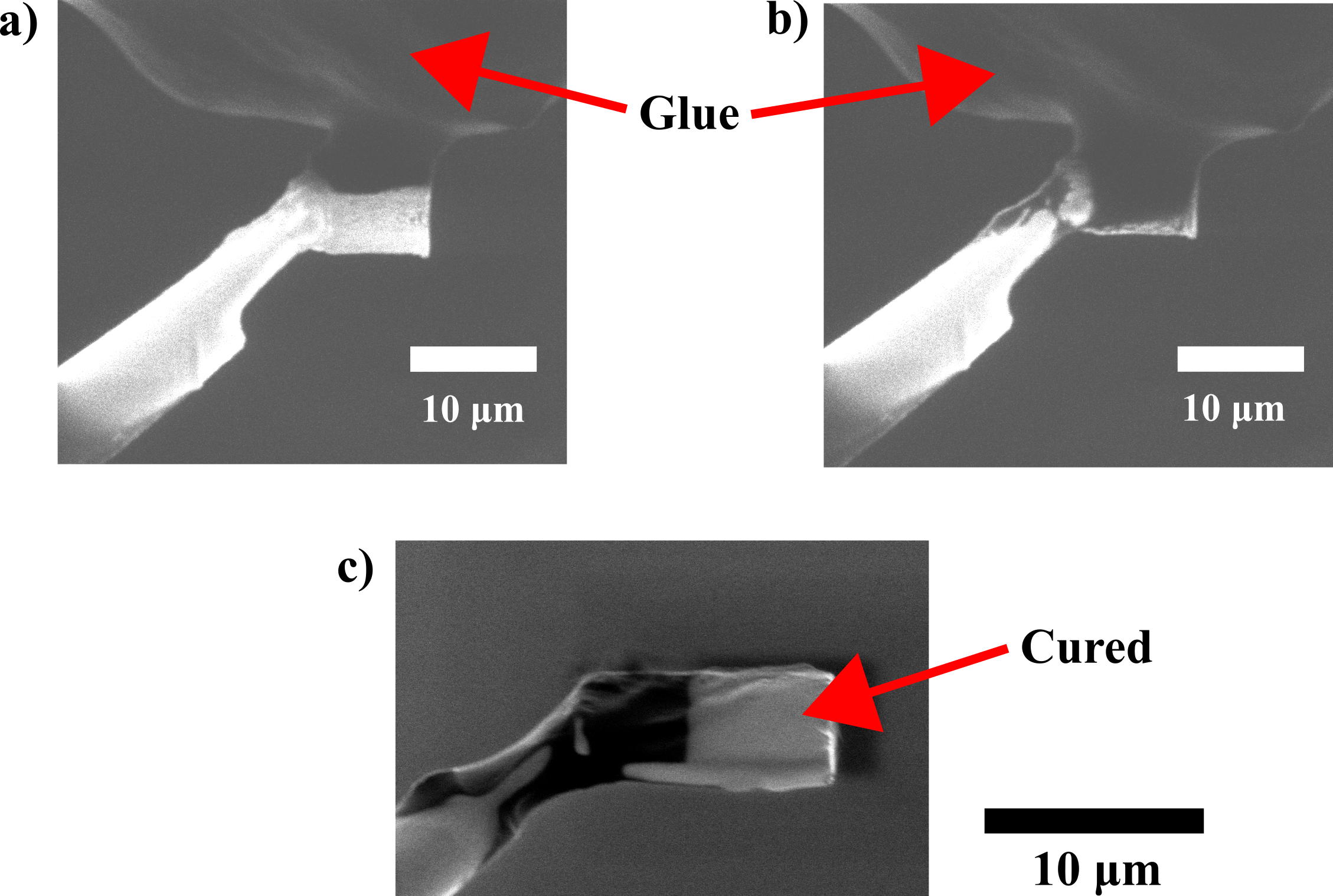}
    \caption{SEM images of the glue filling process showing a) an SnSe lamella on a micromanipulator making contact with a blob of SEMGlu\texttrademark, b) the glue creeping up onto the lamella and into the pores, and c) the coated lamella after part of the glue has been cured}
    \label{fig:GlueCuring}
\end{figure}

To circumvent this problem, and to fully penetrate the pores in a given sample, a post-liftout pore filling method is introduced, using  Kleindiek SEMGlu\texttrademark, a grease-like vacuum-compatible adhesive that hardens under exposure to the electron beam. After liftout with the micromanipulator, the sample is dipped into a blob of SEMGlu\texttrademark placed onto the sample stage. This can be seen in Figure~\ref{fig:GlueCuring}a). The glue is sufficiently pliable that a lamella can be submerged into it and subsequently extracted while remaining on the micromanipulator, yet also viscous enough that a layer will remain on the lamella. Furthermore, the glue's surface tension is sufficiently low that it is pulled up onto the lamella by capillary action, as can be seen in Figure~\ref{fig:GlueCuring}b), which was taken only a few seconds after \ref{fig:GlueCuring}a). This helps not only to wet the lamella with the glue, but also to fill all the pores of the material. 

After extraction from the glue, the lamella is covered with a still pliable layer of glue, which seeps into the pores of the material. The material can then be cured under the electron beam to set it, thus locking the glue into the pores. Figure~\ref{fig:GlueCuring}c) shows the material being cured. The glue can be cured in tens of seconds, and a change in contrast from near-black to a much brighter color signifies that the curing process is complete. Curing takes place through secondary electrons and is best done with the electron beam to reduce damage to the sample. However, in FIB-SEMs without rotatable manipulators the ion beam can also be used, as the secondary electrons generated are still sufficient to cure the material. Because the curing is based on secondary electrons, there is a limit to the maximum thickness of lamellae that can be cured in this manner, though this is voltage and material dependent. Typically, liftout lamellae will be cut to thicknesses of around 1-2\unit{\micro\metre}, and the large area-to-volume ratio makes pores more easily accessible for glue infiltration and facilitates curing all the way through.

After extraction from the glue, the lamella is covered with a still-pliable layer of glue, which seeps into the pores of the material. The material can then be cured under the electron beam to set it, thus locking the glue into the pores. Figure~\ref{fig:GlueCuring}c) shows the material being cured. The glue can be cured within tens of seconds, and a change in contrast from near-black to a much brighter color signifies that the curing process is complete. As the curing relies on secondary electrons, it is most effectively carried out using the electron beam. In FIB-SEMs without rotatable manipulators, however, the ion beam can also be used, as the secondary electrons generated during ion bombardment are sufficient to initiate curing. It should be noted, though, that the use of ions may lead to sample damage — particularly in thin lamellae — and should therefore be used with caution. Even if the electron beam cannot be perfectly aligned, curing with the electron beam at an angle is generally preferable, or alternatively leave your sample thicker than normal and then mill off the ion beam damage at a later thinning stage. Furthermore, subsequent processing steps, such as polishing or final thinning, will likely contribute to completing the curing of any remaining uncured material. However, because the curing process is driven by secondary electrons, there remains a limit to the maximum lamella thickness that can be effectively cured in this way, although this limit depends on both the accelerating voltage and the material. Typically, lift-out lamellae are cut to thicknesses of around 1–2~\unit{\micro\metre}, and their large area-to-volume ratio promotes glue infiltration into the pores and facilitates complete curing throughout the section.

\begin{figure}[h!]
    \centering
    \includegraphics[width=1.0\linewidth]{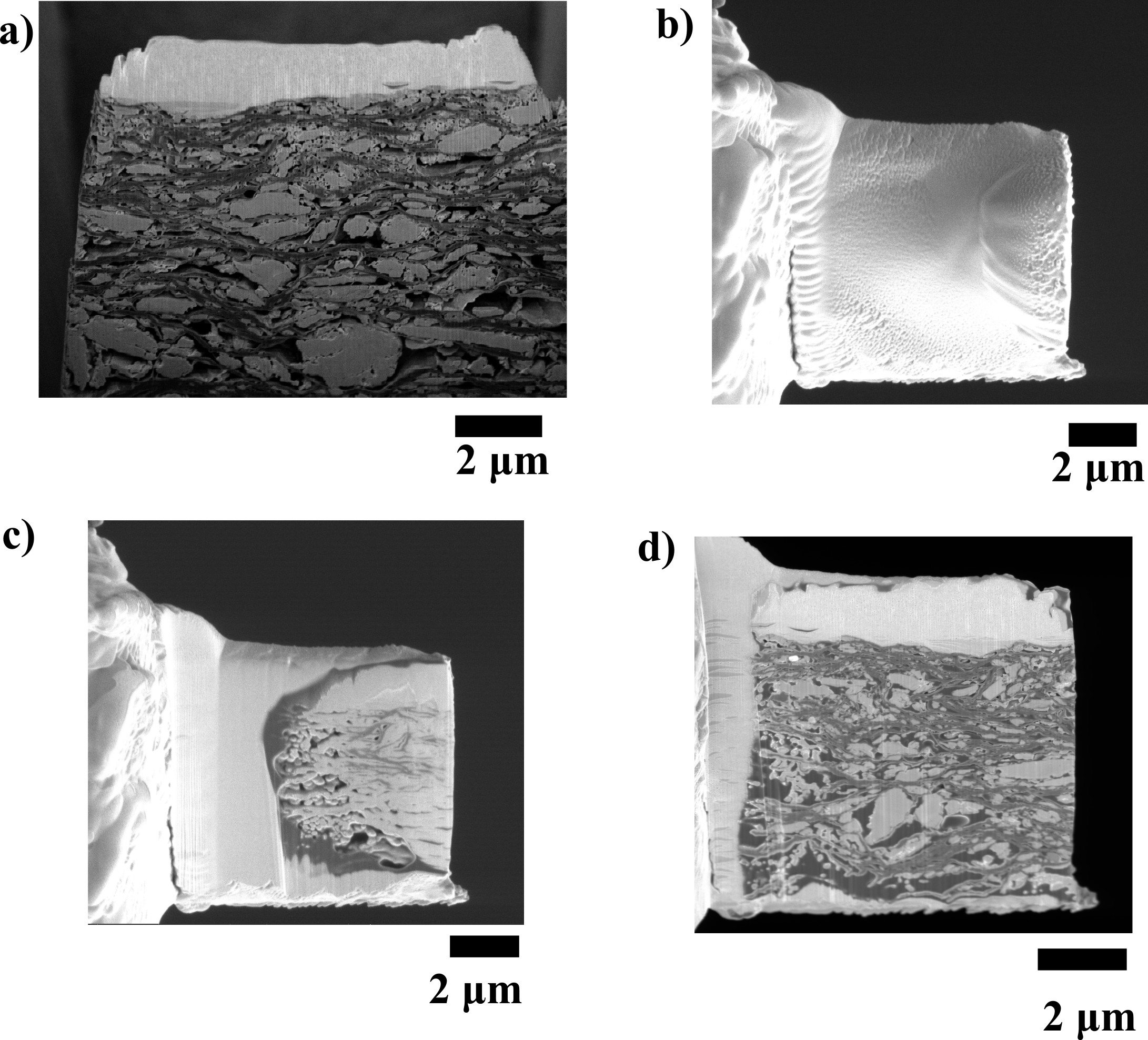}
    \caption{SEM images of an SnSe lamella showing a) the lamella prior to liftout with the micromanipulator, b) the lamella fully coated in cured glue, c) the outer layers of glue being milled away to reveal the lamella underneath, and d) the lamella with the pores now filled almost entirely with cured glue}
    \label{fig:GluedLamella}
\end{figure}

After the glue has been applied and cured, the lamella can then be used like any other for making APT specimens, though for ease of identifying areas of interest, a post-glue polishing step can be useful. This is illustrated in Figure~\ref{fig:GluedLamella}. Figure~\ref{fig:GluedLamella}a) shows a cut lamella of SnSe pre-liftout, with the pores in the structure clearly visible. After the lamella is coated in glue and cured, the underlying structure is covered, as can be seen in Figure~\ref{fig:GluedLamella}b). The top layers of glue can then be removed with the ion beam to reveal the original material and allow for site specific selection during the sharpening process. Figure~\ref{fig:GluedLamella}c) shows the lamella during this process.

Figure~\ref{fig:GluedLamella}d) shows the SnSe lamella after the outer layers of cured glue have been milled away. As can be seen, the glue has filled almost all of the pores in the materials. This now allows for the material to be mounted on Si posts and to be sharpened into APT specimens as in Thompson et al\citep{thompson_situ_2007}. 

\begin{figure}[h!]
    \centering
    \includegraphics[width=0.9\linewidth]{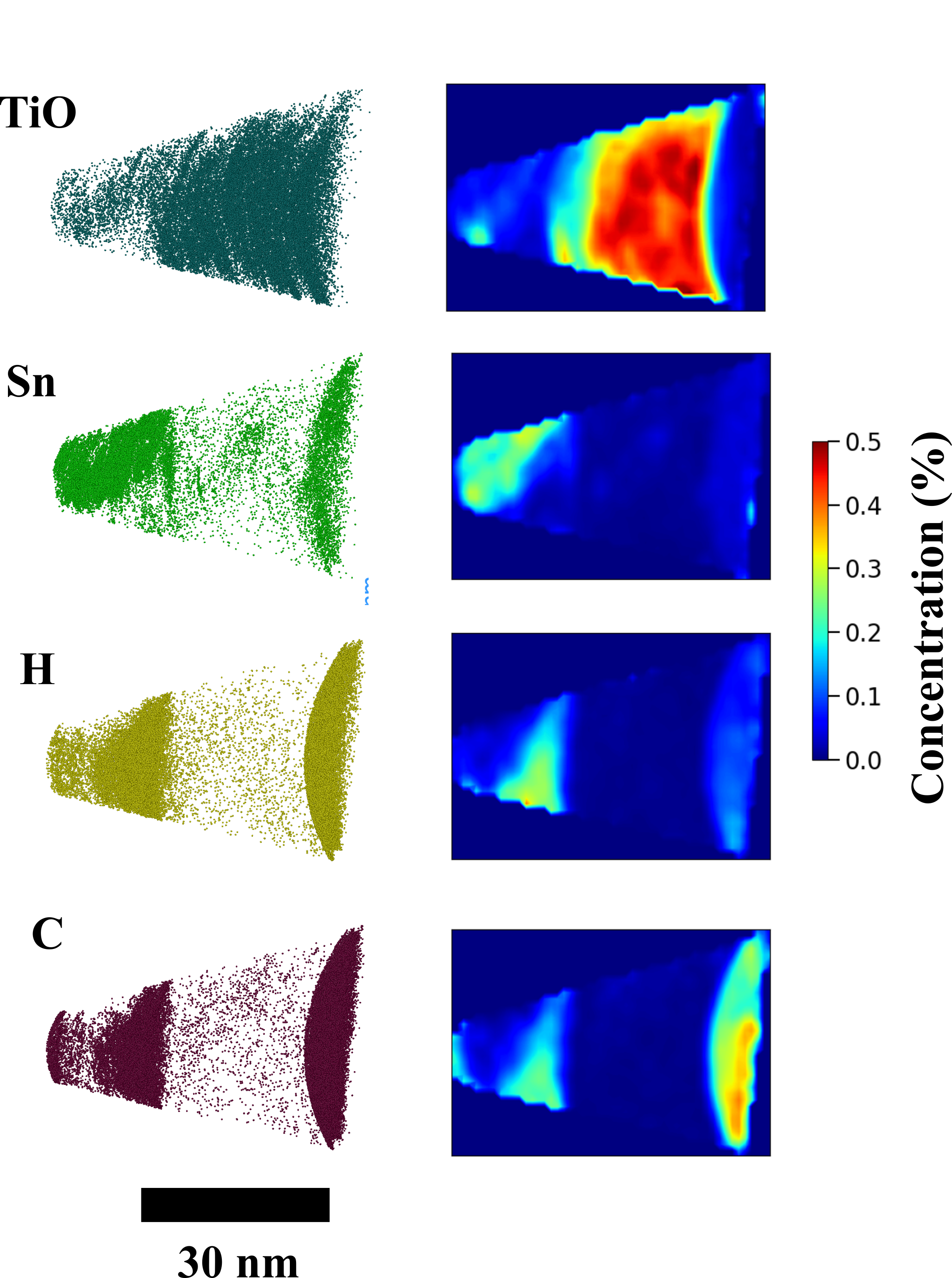}
    \caption{APT reconstruction of an SnSe specimen filled with glue, showing the locations of ions of C, H, Sn, and TiO, as well as the associated concentration heat maps. While the mechanical nature of the mixing process leads to some intermixing, but segregation can nonetheless be observed between the glue (H and C), the particles (Sn), and the Mxene structure (TiO)}
    \label{fig:GlueNeedle}
\end{figure}

Figure~\ref{fig:GlueNeedle} shows an APT reconstruction of a specimen made from a glue-filled lamella of SnSe. As the glue is primarily made up of chains of hydrocarbons, peaks pertaining to $C^{1-2+}$ and $H_{1-3}^{+}$ can be used to track its location within the reconstruction. The particles of SnSe can be visualised by using the  $Sn^+$ spatial distribution. For MXenes, a variety of cluster or molecular ions are detected, $Ti_xO_y^{n+}$ and $Ti_xC_y^{n+}$, as well as various hydrocarbon chains. The ratios between these ions depend heavily on the analysis conditions, but here $TiO^{+}$ was used as the primary ion to image the Mxenes, which is similar to our previous MXene APT paper \cite{kramer_nearatomicscale_2024}. The heatmaps in the figure show the concentrations of these selected ions throughout the reconstructed 3D point cloud.

The reconstruction evidences significant segregation between regions consisting primarily of glue, MXene and SnSe. Due to the mechanical intermixing of the materials during electrode preparation, some spatial overlap of MXene and particle was expected due to mechanical alloying, but there is still visible separation between the two regions. This is likely an area where MXene and particle are touching, with the glue having filled in any remaining gaps.

The advantages of the glue-based preparation method are already apparent in the improved structural integrity of the resulting specimens. Preparation of APT specimens from non-fully dense or weakly cohesive materials without pore filling is typically difficult and unreliable, often leading to samples with compromised geometry, as shown in Figure~\ref{fig:SnSePorosity}. In this context, it should be noted that the challenge is not always related solely to porosity. Materials composed of loosely bound flakes or particles—such as MXene-containing composites, where individual flakes are embedded within a matrix~\cite{kramer_nearatomicscale_2024}, or layered assemblies held together primarily by van der Waals forces~\cite{kramer2024facilitating} can exhibit similar issues. In such systems, weak interfacial bonding and the absence of full densification can result in structural fragility during specimen shaping or field evaporation, even in the absence of true porosity. 

By contrast, preparation of samples with SEMGlu\texttrademark\.  is comparable to the preparation process of solid samples. The only additional steps required are additional thinning of the liftout lamella to expose the relevant pores, dipping into and subsequent curing of the glue, and another thinning step to again expose the material of interest. A practiced FIB operator is capable of performing these steps in a timeframe of around half an hour, and can thus produce sharp specimens with only a small reduction in overall output compared to conventional methods. The only preparation necessary for use of this method is the deposition of a small amount of the glue onto a part of the sample holder accessible with the micromanipulator, requiring no additional instruments or significant preparation time. As this can be easily accomplished in a glovebox, or onto a part of the stage prior to sample preparation commencing, the method is thus compatible with vacuum transfer into the FIB-SEM, and can thus be applied to samples that are sensitive to air or moisture. 

Beyond its structural integrity, both the electrical and thermal conductivities of the specimen can strongly influence the analytical performance of APT. The conductive nature of the cured glue suggests that it could be suitable as a pore-filling material for specimens analysed in voltage-pulsing mode. In laser-pulsing mode, field evaporation is triggered by a transient increase in the specimen’s tip temperature following light absorption. The subsequent cooling occurs via thermal diffusion along the shank, forming a \textit{thermal pulse}. The rate of this cooling is governed by the thermal diffusivity of the specimen, with higher diffusivity leading to shorter cooling times and thus sharper pulses~\citep{vurpillot2006estimation}. Conversely, materials with low thermal diffusivity exhibit slower cooling, which can lead to extended \textit{thermal tails} on mass-spectrum peaks and a reduction in mass resolving power~\citep{camus_simulation_1993, bunton_advances_2007, kelly_laser_2014}. Such tails can make assignment of elemental identities in the mass spectrum significantly more challenging.

There are two distinct aspects that can limit the effective thermal diffusivity of a mounted specimen. The first concerns the internal thermal conduction within the specimen itself, which here consists of a composite of the material, the cured glue, and any remaining porosity. The effective diffusivity in this region is governed by the thermal properties and distribution of these constituents, and it directly affects how rapidly the tip can cool after each laser pulse. The second aspect relates to the thermal conduction pathway between the specimen and its support, typically a commercial Si coupon. Previous work has shown that the relatively low thermal conductivity of SEMGlu\texttrademark\ can lead to pronounced thermal tails in APT data when glue is used to attach a lamella to a substrate prior to sharpening~\citep{rout_adhesive-based_2018}. In the method presented here, the specimen is attached to the support using material deposited from the gas injection system (GIS) as the bonding agent. This metal–carbon composite exhibits a comparatively high thermal conductivity, providing a heat dissipation pathway similar to that of nearly pure metal attachments. As a result, the interface between the sample and the support is not the primary bottleneck for thermal diffusion in this configuration; rather, the limiting factor is likely the composite nature of the specimen itself.

In practice, this means that the impact of the glue on the thermal diffusivity of the overall specimen will depend on both the porosity of specimen, the curing conditions, and the thermal conductivity of the base material. With high porosity, high conductivity samples will likely show a larger drop in overall diffusivity, whereas low porosity, low conductivity samples would be less affected overall. In practice, we observed no difficulties with thermal tails in smaller pores within the sample, but much more significant tails in larger regions of glue, in our case at the bottom of the needle before the sample fractured. A comparison can be found in Figure~\ref{fig:SuppThermalTails} and Table~\ref{tab:SuppMRP}.

In practice, the impact of the glue on the overall thermal diffusivity of the specimen depends on several factors: the porosity of the material, the curing conditions, and the intrinsic thermal conductivity of the base material. The effective heat transport is governed by the relative conductivities and spatial distribution of the glue and the specimen material, such that heat will preferentially travel through the path of least resistance. Consequently, in highly porous but intrinsically conductive materials, the introduction of low-conductivity glue can lead to a pronounced reduction in effective diffusivity. Conversely, for dense materials with low intrinsic conductivity, the relative impact of the glue may be less severe—but it can still remain significant if the glue dominates the thermal pathway. In our experiments, we observed no apparent difficulties with thermal tails associated with small pores infiltrated by glue, but more pronounced tails when larger glue regions were present, particularly near the bottom of the needle, just before fracture. A comparison is shown in Figure~\ref{fig:SuppThermalTails}.

SEMGlu\texttrademark has also been demonstrated to work as an adhesive at cryogenic temperatures, with the glue eventually freezing solid but maintaining the ability to cure under the electron beam\citep{mulcahy_look_2025}. This opens up the possibility of using this method for beam-sensitive samples, with glue infiltration done at room temperature and thinning and possible curing steps performed at cryogenic temperatures to limit beam damage.

\section*{Conclusion}

In conclusion, the efficacy of a SEMGlu\texttrademark-based pore filling method for APT analysis has been successfully demonstrated. In contrast to previously reported approaches, it is easily accessible without further instrumentation or preparatory steps, and can be carried out in vacuum directly in the SEM-FIB. The necessary structural integrity required for APT specimen preparation is provided by the glue, which also provides electrical and thermal conduction across the different parts of the material and facilitate contact to the substrate. In some samples, thermal tails may be a concern, though they seem to only be prominent in areas with large amounts of glue.
\clearpage

%%%%%%%%%%%%%%

\section*{Supporting Information}

\renewcommand{\thefigure}{S\arabic{figure}}
\setcounter{figure}{0}

\begin{figure}[h]
    \centering
    \includegraphics[width=1.0\linewidth]{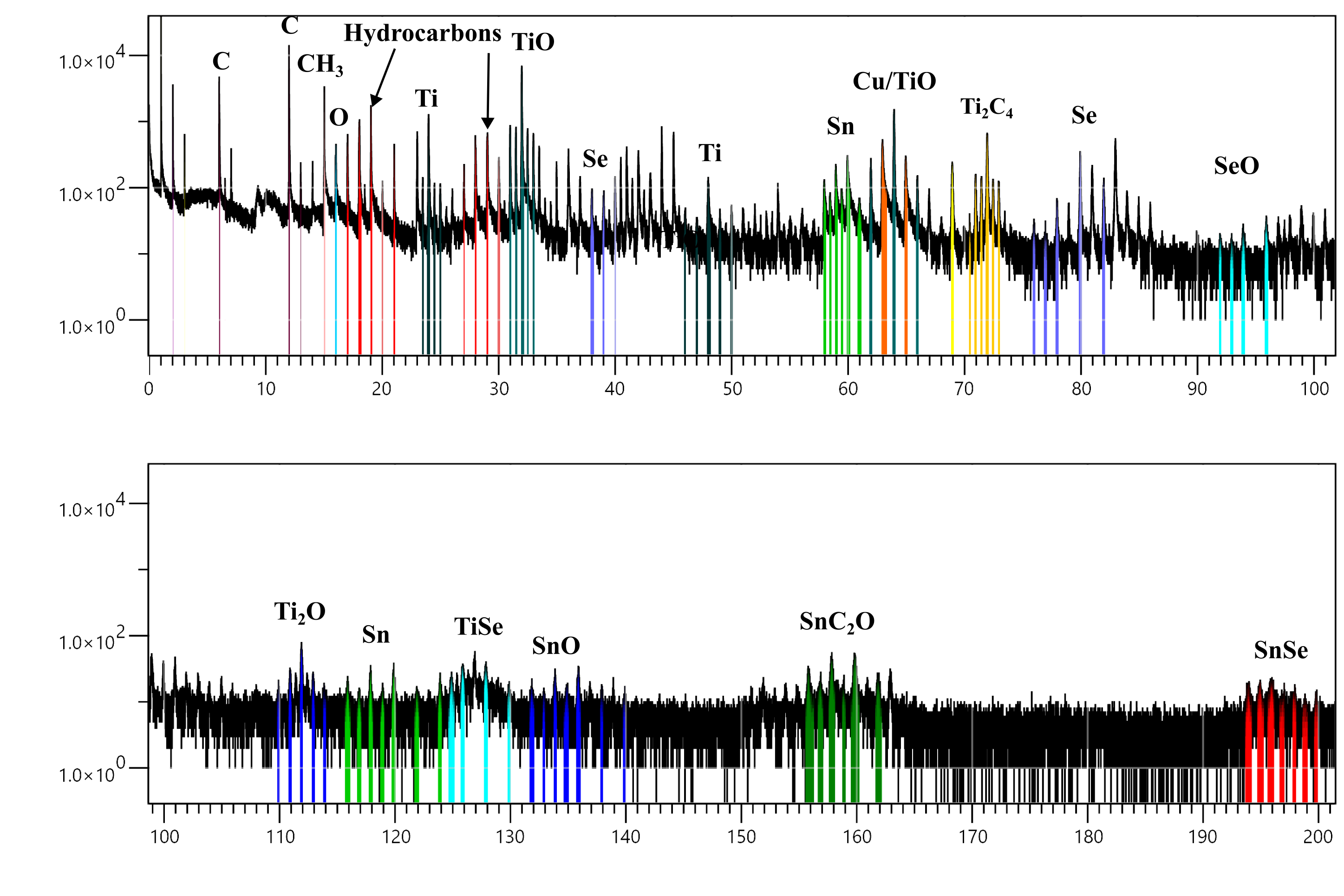}
    \caption{Mass-to-Charge spectrum of APT analysis shown in figure \ref{fig:GlueNeedle}}
    \label{fig:SuppAPTSpectra}
\end{figure}

The spectrum of the APT analysis is shown in Figure~\ref{fig:SuppAPTSpectra}. The hydrocarbons in the glue, as well as the carbon present in the MXenes in combination with oxidation, produce a large number of peaks in the final spectrum, and make ranging of all peaks difficult. However, peaks associated with the main components of the sample (Ti, TiO, Sn, Se, etc.) are still very prominent and easy to identify, allowing for analysis of the sample. More details on the unranged ions are discussed in Table~\ref{tab:SuppAPTComposition}
\clearpage

\begin{figure}[h]
    \centering
    \includegraphics[width=\linewidth]{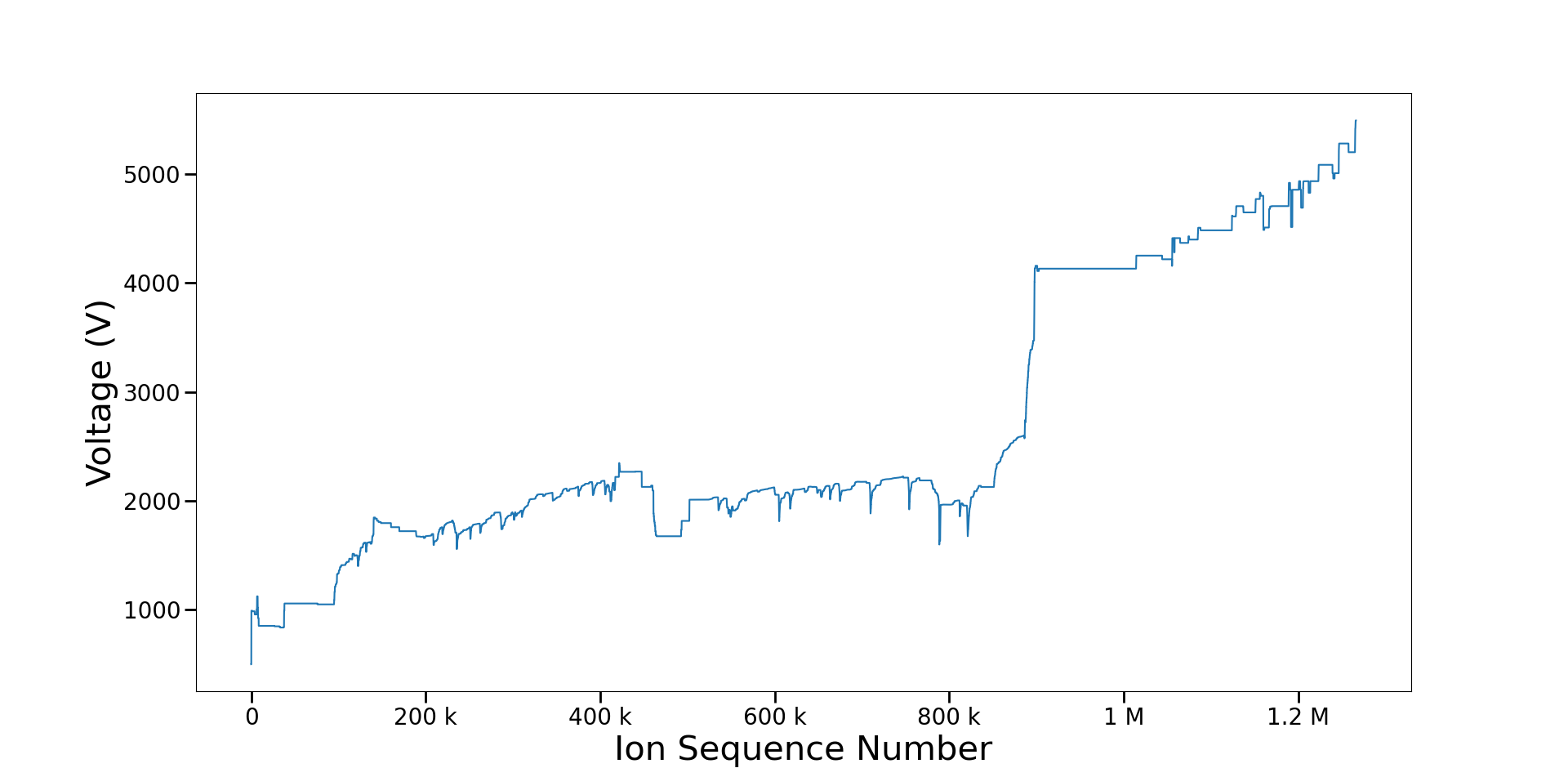}
    \caption{Voltage curve of the the APT analysis shown in figure \ref{fig:GlueNeedle}}
    \label{fig:SuppVoltage}
\end{figure}
\clearpage

\begin{figure}[h]
    \centering
    \includegraphics[width=0.8\linewidth]{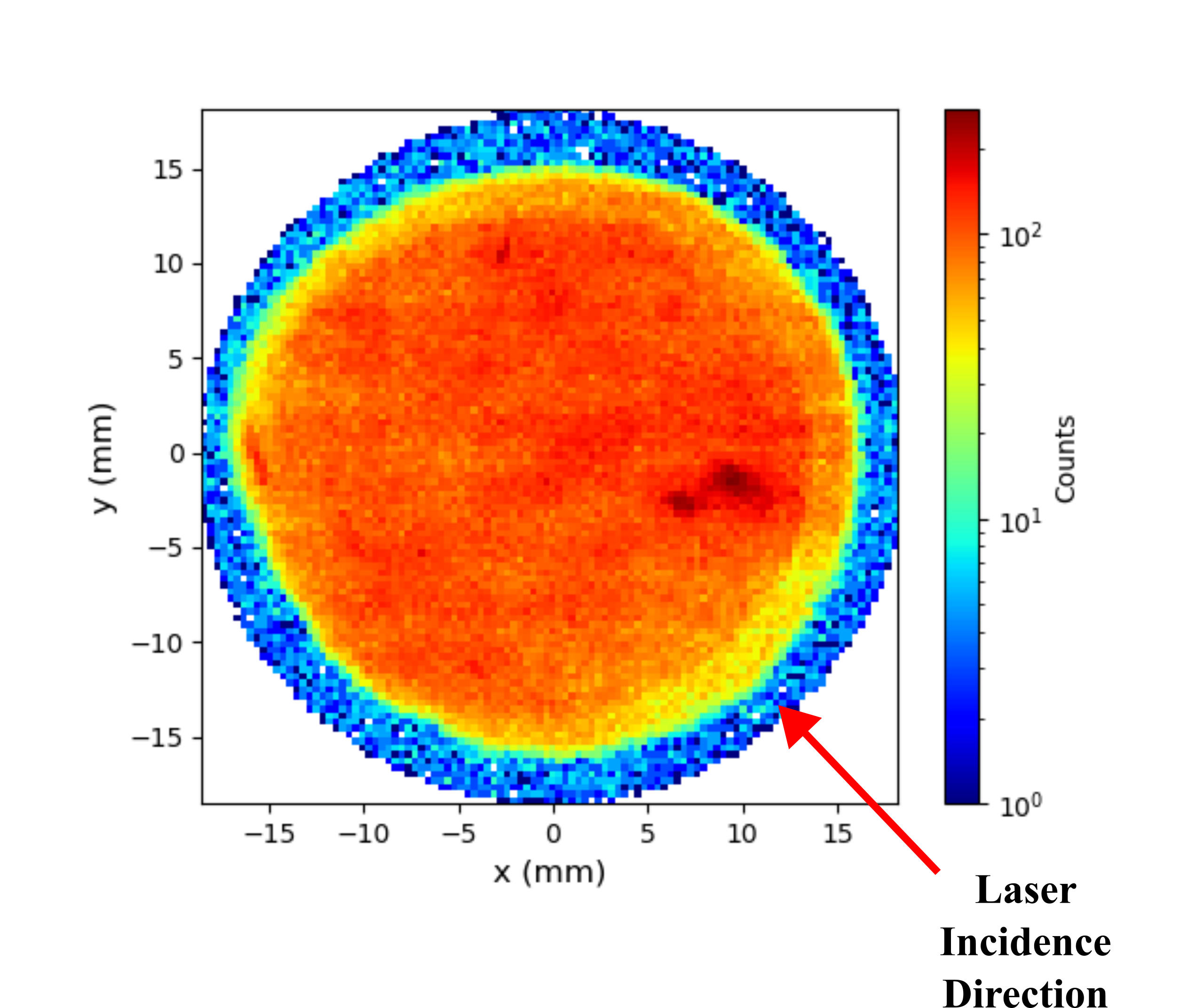}
    \caption{Detector map of the APT analysis shown in figure \ref{fig:GlueNeedle}}
    \label{fig:SuppAPTDetector}
\end{figure}
\clearpage

\begin{table}[h]
    \centering
    \begin{tabular}{c|c}
        Ion & Compositional percentage \\ \hline
        TiO & 20.32\\
        Unranged & 17.76\\
        C & 11.19\\
        Hydrocarbons & 9.66\\
        H & 6.35\\
        Sn & 4.95\\
        Ti & 4.19\\
        Cu & 3.88\\
        Ti$_2$C$_4$ & 3.65\\
        SeH$_3$ & 3.02\\
        Se & 2.89\\
        CH$_3$ & 2.30
    \end{tabular}
    \caption{Percentage presence of all ions \textgreater 2\% in the APT analysis}
    \label{tab:SuppAPTComposition}
\end{table}

Table~\ref{tab:SuppAPTComposition} shows a breakdown of the composition of major ions in the APT analysis. The percentage of unranged ions is significant at around 18\%. However, the biggest contributors to this are the very top of the needle, where the unranged ions are briefly \textgreater 37\%, and the MXene, where around 20\% of ions are unranged, the latter region being a significant portion of the specimen. While APT studies on 2D materials have so far been uncommon, the study done by Kr\"amer et al. in 2024 supports the complexity of the MXene spectrum, leading to significant difficulties in ranging \cite{kramer_nearatomicscale_2024}. In other areas of interest, such as the SnSe particle and the glue, this percentage is lower, at around 12\%. This is still significant, but should allow for reconstruction of these areas with some accuracy. In addition, MXene components are still found in these regions, though in smaller percentages, and may be contributing to this unranged percentage. Materials that produce more easily ranged spectra are likely to show a much less significant proportion of unranged ions when used with the glue method discussed here. The general lack of APT analyses on 2D materials such as MXenes also shows the overall difficulty in analysing these materials in the atom probe, a problem that the pore filling method shown here presents a possible solution for.
\clearpage

\begin{figure}
    \centering
    \includegraphics[width=0.8\linewidth]{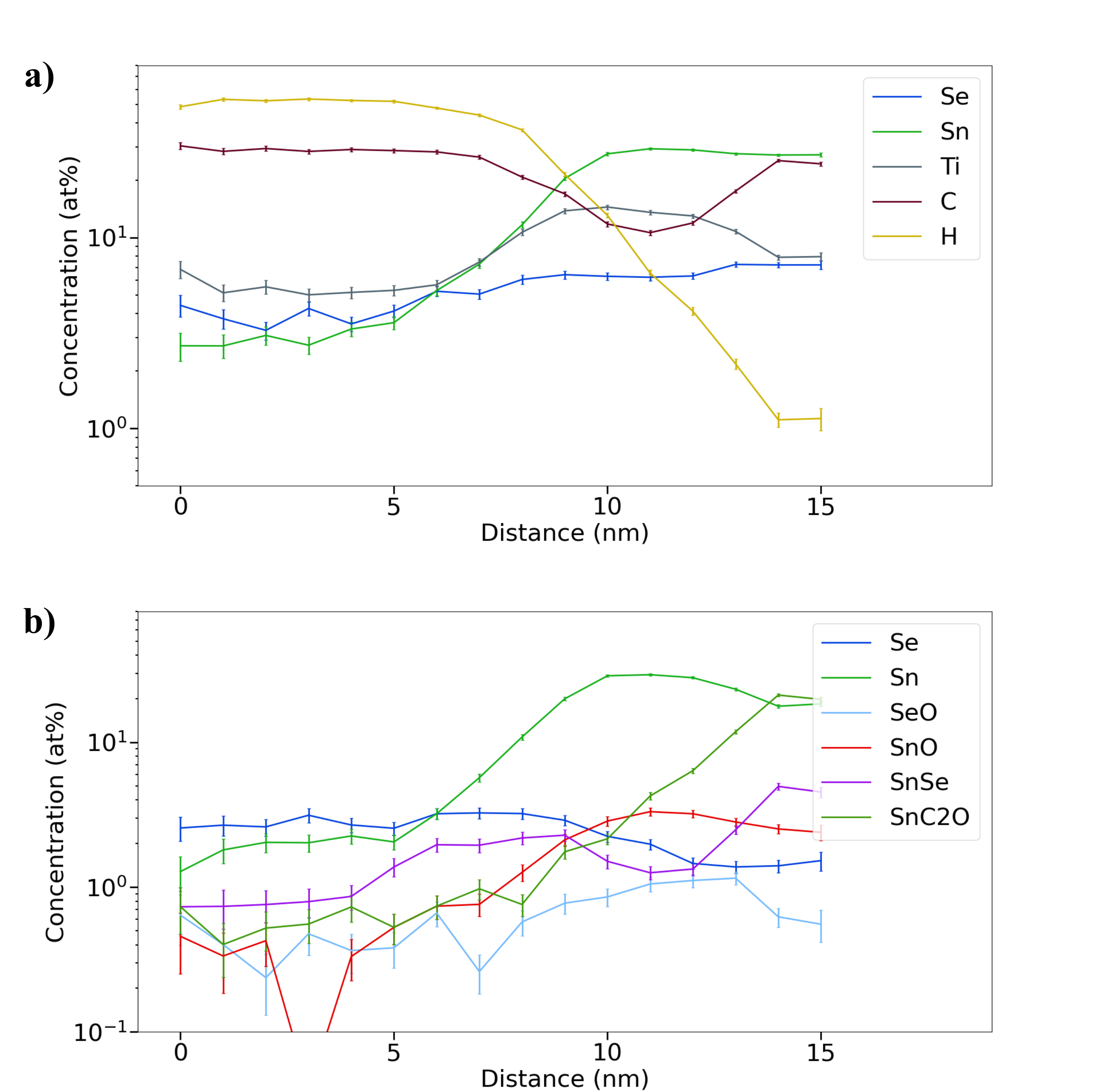}
    \caption{1D concentrarion profile along the interface between the glue and the SnSe particle, showing a) the decomposed atoms and b) the individual ion species making up the Sn and Se components}
    \label{fig:SuppSnSeGlueInterface}
\end{figure}

Figure~\ref{fig:SuppSnSeGlueInterface} shows the 1D concentration of decomposed elements at the interface. As can be seen in Figure~\ref{fig:SuppSnSeGlueInterface}a), the interface shows a significant rise in Sn and a smaller but still noticeable rise in Se coming into the SnSe particle. The levels of both then remain relatively constant throughout the particle. Figure~\ref{fig:SuppSnSeGlueInterface}b) further shows the individual ions making up the Sn and Se concentrations. Here, it can be seen that there is a region of relative stability close to the interface, followed by significant shifts going further in, especially a sharp rise in SnC$_2$O. However, the particle here interfaces directly with the edge of the needle, so this change may be due to those effects. In either case, the interface with the glue shows relative stability in composition. 
\clearpage

\begin{figure}[h]
    \centering
    \includegraphics[width=1.0\linewidth]{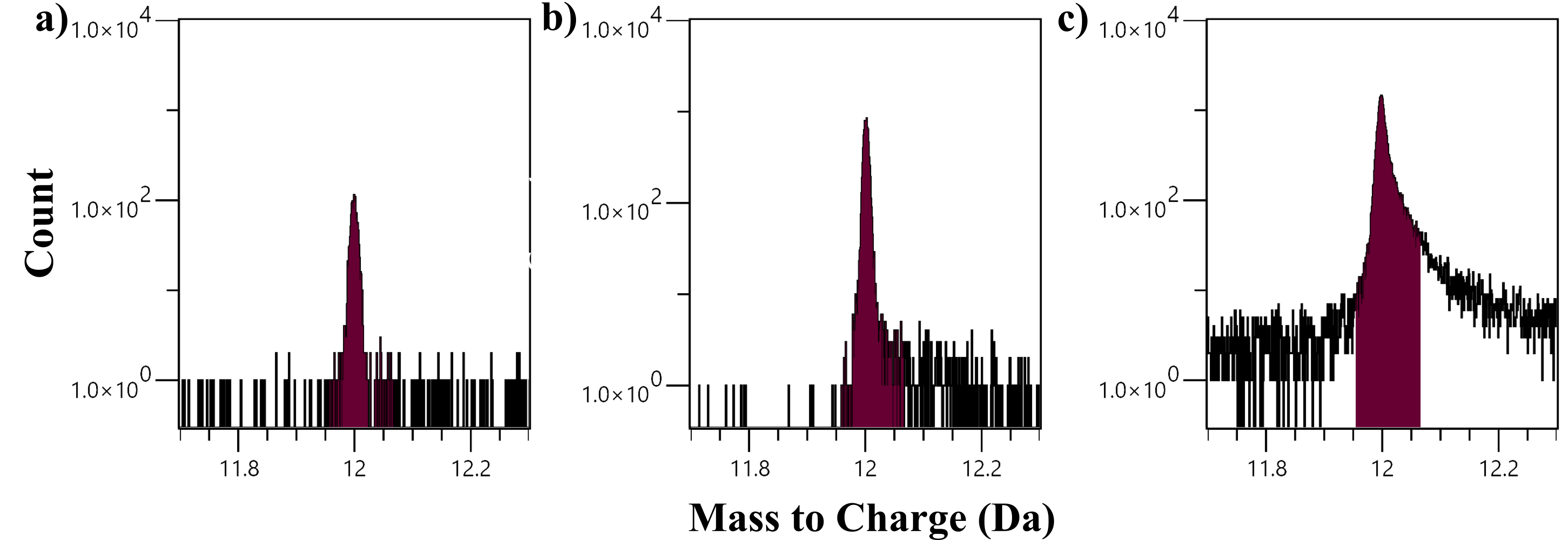}
    \caption{Comparison of thermal tails in the APT analysis in the region of a) the SnSe particle, b) the small glue pore near the particle, and c) the region of glue at the bottom of the needle. Thermal tails are minimal in both the particle and glue pore, but much more pronounced in the glue at the bottom of the needle. Resolving power of these peaks is shown in Table~\ref{fig:SuppThermalTails}}
    \label{fig:SuppThermalTails}
\end{figure}
\clearpage

\begin{table}[h]
    \centering
    \begin{tabular}{c|ccc}
        Peak & FWHM & FWTM & FW100M \\ \hline
        SnSe Particle & 980 & 505 & 318\\
        Glue Pore & 1280 & 668 & 396 \\
        Bottom of Needle & 952 & 307 & 104
    \end{tabular}
    \caption{Mass resolving power (MRP) of the Carbon-12 peaks shown in Figure~\ref{fig:SuppThermalTails}, given at full width half maximum (FWHM), full width tenth maximum (FWTM), and full width hundredth maximum (FW100M). Significant loss of resolving power due to thermal tails can be seen at the bottom of the sample, especially in the FW100M}
    \label{tab:SuppMRP}
\end{table}
\clearpage

\section*{Competing interests}
No competing interest is declared.

\section*{Author contributions statement}

M.S.C. and J. O. D. conceived the idea of using SEMGlu™ to prepare porous samples for APT. L.W. and M.S.C. conducted the SEM and FIB experiments. L.W., J.O.D. and M.S.C. analyzed the APT specimens and processed the data. All authors discussed the results and contributed to the final version of the manuscript.

\section*{Acknowledgments}
This work was made possible by the EPSRC Cryo-Enabled Multi-microscopy for Nanoscale Analysis in the Engineering and Physical Sciences EP/V007661/1. L.W. and K.A. acknowledged the EPSRC Centre for Doctoral Training in the Advanced Characterisation of Materials (CDTACM)(EP/S023259/1) for their PhD studentship funding.  M.S.C. acknowledges funding from Royal Society Tata University Research Fellowship (URF\textbackslash R1\textbackslash 201318), Royal Society Enhancement Award\\ (RF\textbackslash ERE\textbackslash 210200EM1) and ERC CoG DISCO grant 101171966. B.G.
acknowledges financial support from the ERC-CoG-SHINE-771602. We thank Teng Zhang and Yury Gogotsi from Drexel University for their Mxene samples. 

\bibliographystyle{unsrtnat}
\bibliography{main}
%\bibliography{reference}
\end{document}